\documentclass[fleqn,10pt]{wlscirep}
 \pdfoutput=1 
\usepackage[utf8]{inputenc}
\usepackage{amsmath}
\usepackage[T1]{fontenc}
\title{General scaling in bidirectional flows of self-avoiding agents}

\author[1,*]{Javier Crist\'{i}n}
\author{Vicen\c{c} M\'{e}ndez}
\author{Daniel Campos}
\affil{Grup de F\'{i}sica Estad\'{i}stica. Departament de F\'{i}sica. Universitat Aut\` {o}noma de Barcelona. \\ 08193 Bellaterra (Barcelona) Spain}

\affil[*]{javier.cristin@uab.cat}

\begin{abstract}
The analysis of the classical radial distribution function of a system provides a possible procedure for uncovering interaction rules between individuals out of collective movement patterns. A formal extension of this approach has revealed recently the existence of a universal scaling in the collective spatial patterns of pedestrians, characterized by an effective potential of interaction $V(\tau)$ conveniently defined in the space of the times-to-collision $\tau$ between the individuals. Here we significantly extend and clarify this idea by exploring numerically the emergence of that scaling for different scenarios. In particular, we compare  the results of bidirectional flows when completely different rules of self-avoidance between individuals are assumed (from physical-like repulsive potentials to standard heuristic rules commonly used to reproduce pedestrians dynamics). We prove that all the situations lead to a common scaling in the $\tau$-space both in the disordered phase ($V(\tau) \sim \tau^{-2}$) and in the lane-formation regime ($V(\tau) \sim \tau^{-1}$), independent of the nature of the interactions considered. Our results thus suggest that these scalings cannot be interpreted as a proxy for how interactions between pedestrians actually occur, but they rather represent a common feature for bidirectional flows of self-avoiding agents.
\end{abstract}
\begin{document}

\flushbottom
\maketitle
%
%
\thispagestyle{empty}

\section*{Introduction}

The theoretical description of pedestrian flows represents a field of the greatest importance due to its direct impact on issues related to urban planning, monitoring of public spaces or optimization of evacuation protocols, to name a few \cite{helbing01,johansson09,zheng12,kirchner02}. In the last years, facilitated by an improvement in simulation capacities and in the availability of experimental data, this topic has attracted physicists for its interest as a multiagent system driven by non-trivial rules of ordering, alignment and self-avoidance, among other. Thus, a significant effort has been put both in (i) understanding these interaction rules in order to recover the patterns experimentally observed \cite{degond13,rahmati18} and (ii) identifying the minimal or toy models which are able to capture the essentials of such phenomena \cite{helbing00,goldsztein17}.

Lane or trail formation in crowds, in particular, has been extensively studied as a manifestation of self-organization in pedestrian flows, both theoretically \cite{yu05,chraibi10,burger16,marschler16,guo18} and through controlled experiments \cite{zhang12,feliciani16}. While in the case of humans one could attribute lane formation to intelligent and efficient decision-making  based on visual information and subsequent prospection, such patterns have been shown to arise even in extremely simple models of agent interaction. The Vicsek model for swarming dynamics \cite{vicsek95,ma13} is probably the best known, but many variations and generalizations based on lattice-gas \cite{muramatsu99,suma12} or \textit{social force} \cite{helbing95,helbing02} models have been developed. Additionally, the formation and sustainance of bidirectional (trail) flows represent also an intriguing situation of interest in behavioral biology too, since only a few social species in the animal kingdom are able to exhibit such behavior in natural conditions \cite{fourcassie10}, ants being the most renowned case \cite{john04,bouchebti15}.

Despite all this multidisciplinary interest, the heterogeneity of models used nowadays to generate such flows/dynamics sometimes goes against the possibility of finding general and far-reaching conclusions. So, existing models/works can sometimes provide different or even contradictory conclusions \cite{feliciani16}. Works aimed at providing unified frameworks and/or at revealing the common properties of these approaches are then convenient to promote understanding and theoretical research within the field \cite{martin18}. 

Within this context, a valuable insight has been recently provided by Karamaouzas et. al. \cite{karamouzas14}. By analyzing 1500 trajectories of pedestrians in outdoor environments they found consistent evidence for an effective potential of interaction $V(\tau)$ between pairs, which was found to depend only on the time-to-collision $\tau$ between the individuals and not on the interparticle distance as in classical fluid systems. This is in line with recent approaches that introduce prospection of future outcomes as the main driving force for intelligent agents, as those based on causal entropic forces \cite{wissner13,mann15}. The analysis carried out in \cite{karamouzas14} yielded $V(\tau) \sim \tau^{-2}$, at least in the significant range of scales where self-avoidance between the pedestrians is relevant. So, a mechanical (Langevin-like) model based on this interaction potential could adequately reproduce the dynamics of pedestrians in real scenarios.

Intrigued by these results, we use here numerical simulations to explore the convenience of a description based on the $\tau$-space for different bidirectional systems based on radically different rules of self-avoidance. 
In general, bidirectional flows emerge from the tension between pair interactions and the existence of two subpopulations that have preference for moving in opposite directions; when the density of individuals $\rho$ is large enough the freedom of movement gets reduced and interactions will be then largely governed by self-avoidance rules. 

\section*{Agent based model}

We consider in the following a multiagent system where the dynamics of the (identical disk-like) agents are governed by a force
\begin{equation}
 \mathbf{F}= \xi \left( \mathbf{v}^{(pr)}-\mathbf{v} \right) + \mathbf{F}^{(sa)}.
\label{eq1}
\end{equation} 

Here, the first term accounts for the preference of each individual to maintain its preferred velocity $\mathbf{v}^{(pr)}$ (denoting $\mathbf{v}$ as the actual velocity) with a certain intensity that we call the \textit{stubbornness}, $\xi$. In order to generate bidirectional fluxes we assume two different subpopulations with the same number of agents each, whose preferred velocities are the same in modulus but have opposite directions. In particular, to avoid spuriors effects in the simulations we sample for each agent a stochastic preferred speed from a Gaussian distribution with mean $\langle  v^{(pr)} \rangle =1.3$ m/s and standard deviation $ \sigma_v = 0.1$ m/s. These specific values are in agreement with those used in \cite{karamouzas14} and similar works on pedestrian dynamics \cite{helbing05,huber14}. Anyway, we have numerically checked that the results reported below are independent of the specific values chosen.

\begin{figure}[h!]
	\centering
	\includegraphics[width=0.65\linewidth]{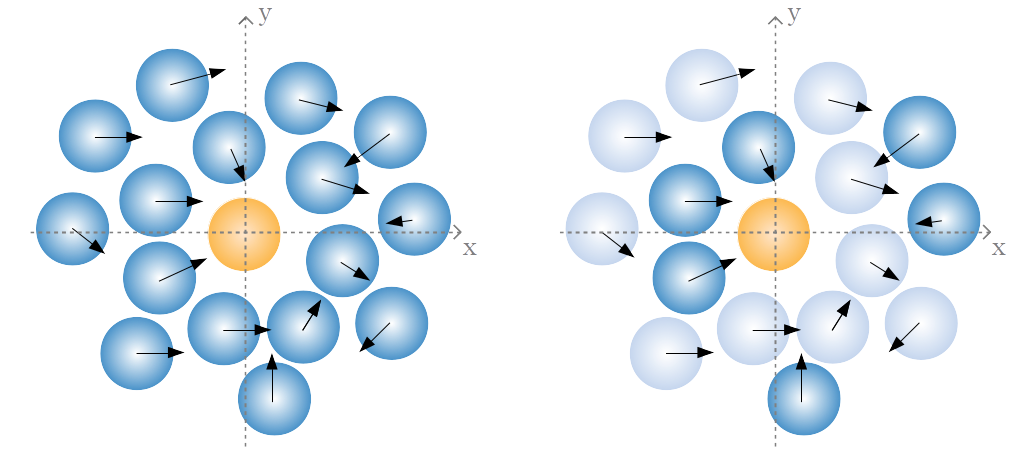}
	
	\caption{\textit{Representation of the differences between the interactions based on the distance $r$ (left) and those in the $\tau$-space (right). The arrows represent the relative velocity of the agents respect to the orange agent at the origin. The individuals filled with solid blue are the only ones contributing to interactions with the orange one (it is, in the second case only those for which a finite and positive $\tau$ can be defined).}}
	\label{fig:radialscheme}
\end{figure}

On the other hand, $\mathbf{F}^{(sa)}$ stands for the pair (self-avoiding) interaction between agents. We here compare the results for three rules/mechanisms based on completely different grounds. The first one consists of a classical pair repulsive interaction in the radial direction, $F_{\textbf{rep}}^{(sa)} \sim r^{-k}$ (where $r$ is the distance between pairs of individuals), with $k>0$. The second one corresponds to the effective potential empirically obtained in \cite{karamouzas14}, this is, a repulsion in the time-to-collision (ttc) space, $F_{\textbf{ttc}}^{(sa)} = -\nabla V_{\textbf{ttc}}$, with $V_{\textbf{ttc}} \sim \tau^{-2}$. These two mechanisms then are reminiscent of typical interaction potentials from statistical mechanics, though in the second case we are considering that the relevant space for interactions is that of $\tau$, so assuming that our intelligent agents can somehow \textit{prospect} future collisions and adapt its behavior to the outcome of such prospections. So, the space of interactions is defined only for those individuals for which $\tau$ is finite, this is, for the set of pairs that will collide at some future time provided its present velocity is mantained (see Fig. \ref{fig:radialscheme}). Finally, as a third case, $F_{\text{heu}}^{(sa)}$, we consider a nonphysical (heuristic) rule which has been found to reproduce most features (e.g stop-and-go waves, turbulent dynamics,...) of collective behavior in pedestrians \cite{moussaid11}. This rule consists of recomputing continually the direction of motion in order to maximize at each step the distance that the agent could travel without colliding with other agents (see the Methods Section for the finer details for the implementation of the self-avoiding rules).

Numerical implementation of our multiagent system identifies, as expected, the existence of a phase transition from a disordered state to lane formation as a function of the values of $\rho$ and $\xi$. To characterize this transition we use the order parameter
\begin{equation}
\phi=\langle  \cos( \theta) \rangle,
\label{eq:op}
\end{equation}
where $\theta$ is the angle between the actual and the preferred velocities (this is, $\mathbf{v}$ and $\mathbf{v}^{(pr)}$) and the average is carried out over all the agents in the system. So, $\phi \to 0$ corresponds to the disordered state in which individuals cannot follow its preferred direction and spend their time avoiding collisions in all directions, while $\phi \to 1$ represents the case where the agents are able to follow its desired direction of motion by adopting a collective pattern with alternate lanes in one direction and the other.

 \begin{figure}[h!]
	\centering
	\includegraphics[width=0.65\linewidth]{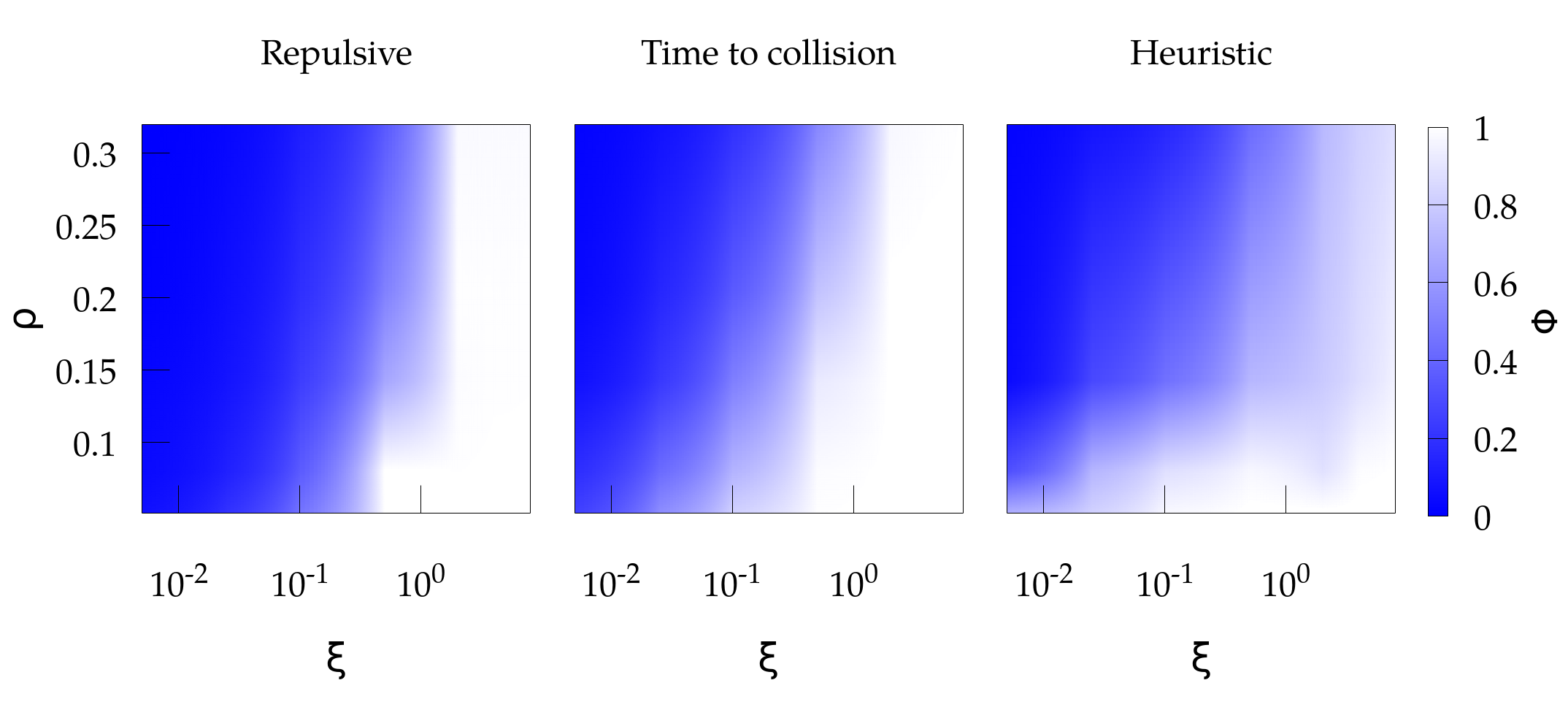}
	\caption{\textit{Phase diagram of the order parameter $\phi$ for the three different mechanisms of self-avoidance as a function of the density of the system $\rho$ and the \textit{stubbornness} $\xi$. The repulsive interaction is fixed to $k=4$.}}
	\label{fig:figop}
\end{figure}

\section*{Results and discussion}

\begin{figure}[h!]
	\centering
	\includegraphics[width=0.65\linewidth]{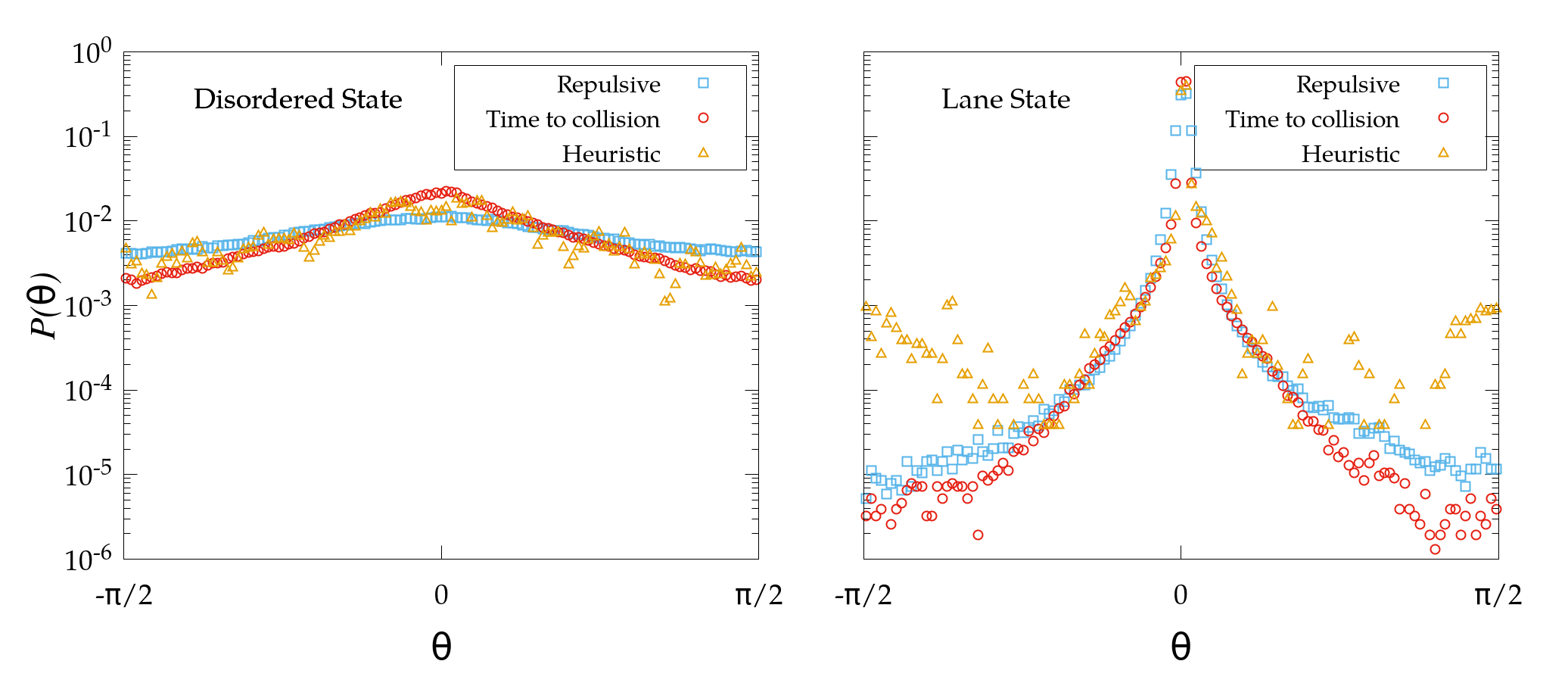}
	\caption{\textit{Probability distribution of the relative angle $\theta$. The left image corresponds to $\xi=0.025$ and the right image for $\xi=2$, with $\rho=0.14$ in both cases. The blue squares correspond to the repulsive interaction ($k=4$), the red circles correspond to the time to collision interaction, and the orange triangles correspond to the heuristic rule.}}
	\label{fig:ang}
\end{figure}

Figure \ref{fig:figop} confirms that the three mechanisms of self-avoidance exhibit qualitatively a very similar behavior for the order parameter $\phi$ at the stationary state. Low stubbornness and high densities promote the disordered phase, while for high stubbornness and low densities the system self-organizes into a new phase where lanes are formed (we provide in the Supplementary Information short videos showing the dynamics observed in each of these two phases for the three self-avoidance mechanisms mentioned above). 
In the disordered state the difference between the actual direction of motion and the preferred one is relatively homogeneous in $(0, \pi)$. We can plot the probability distribution $p(\theta)$ of angle $\theta$ to visualize this (Fig. \ref{fig:ang}, left).  For the case of lanes, on the contrary, most of the individuals move in their desired direction and then the probability distribution becomes clearly peaked at $\theta=0$. Additionally, we observe how the heuristic mechanism exhibits a larger probability  for large deviations in the lane state than the other interactions; this is due to the intrinsic properties of the algorithm, which allows larger reorientations provided they satisfy the maximization of the traveled distance, as explained above.

\begin{figure}[h!]
  \centering
  \includegraphics[width=0.65\linewidth]{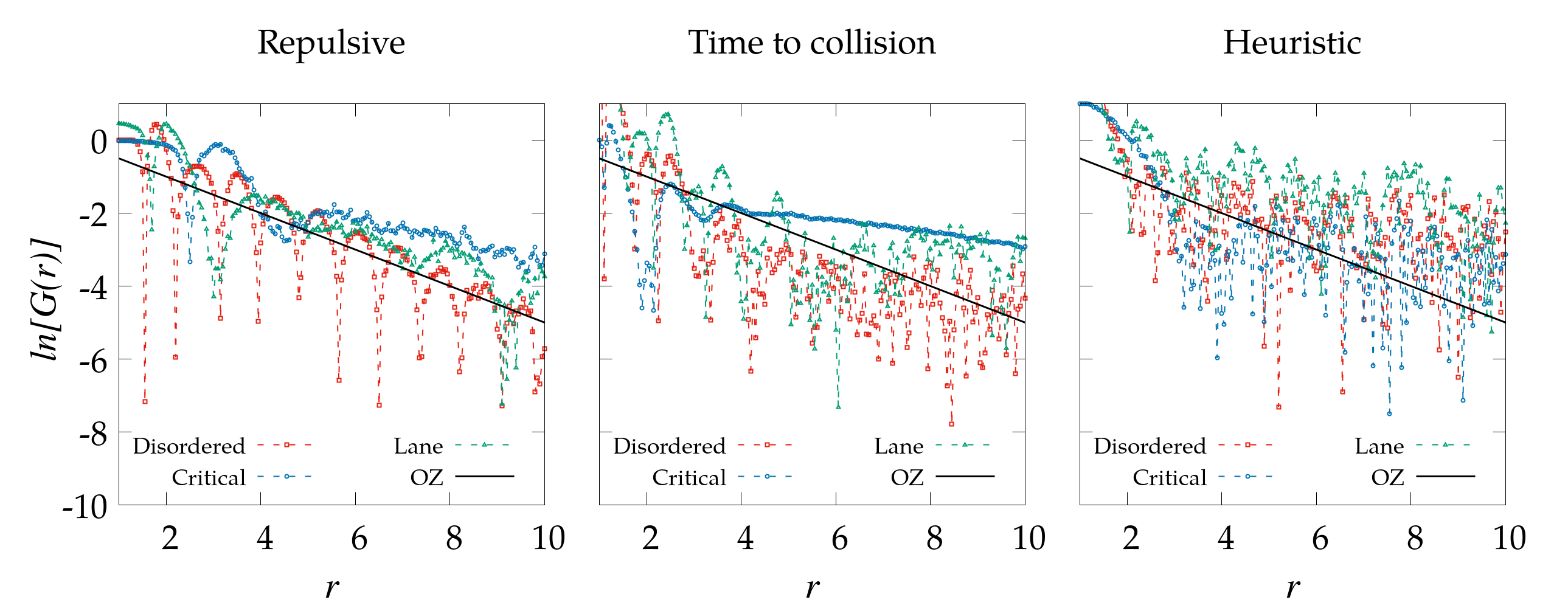}

    \caption{\textit{Comparison between  $\ln[G(r)]$ as a function of the radial distance $r$ over the phase diagram. The black line corresponds to the OZ approximation ($\sim \ln{G(r)} \sim -0.5 \ln{r}$) which is introduced for visual comparison. The repulsive potential ($k=4$) curves (left) correspond to a) $\rho=0.32$ and $\xi=0.025$ (disordered phase), b) $\rho=0.05$ and $\xi=0.5$ (critical region) and c) $\rho=0.14$ and $\xi=2$ (lanes phase). The time to collision curves (center) correspond to a) $\rho=0.32$ and $\xi=0.025$ (disordered phase), b) $\rho=0.32$ and $\xi=0.1$ (critical region) and c) $\rho=0.32$ and $\xi=4$ (lanes phase). The heuristic curves (right) correspond to a) $\rho=0.14$ and $\xi=0.025$ (disordered phase), b) $\rho=0.08$ and $\xi=0.5$ (critical region) and c) $\rho=0.14$ and $\xi=4$ (lanes phase). }}
  \label{fig:OZ}
\end{figure}

To understand in greater detail the properties of these two phases, we next study the spatial distribution of the agents in stationary conditions through the radial distribution function $g(r)$ from the simulations carried out using the three self-avoiding mechanisms mentioned above. As in classical fluids, $g(r)$ here compares the density of interacting agents at a distance $r$ with the density obtained for a non-interacting system, with $g(r) \to 1$ as $r \to \infty$. The corresponding results are presented in Fig. \ref{fig:OZ}. Despite some qualitative differences found due to the different nature of the self-avoiding mechanisms, we observe that the results are relatively consistent with those from the classical Ornstein-Zernike (OZ) approximation for fluid systems \cite{stanley71}, which predicts an asymptotic decay $G(r) \to r^{-0.5}$ (with $G(r) \equiv g(r)-1 $) when the system is far from the critical region where the phase transition occurs. The oscillatory behavior observed there is also characteristic from similar statistical analysis on fluids \cite{savenko05}. All the points of the phase diagram that are far from the critical region satisfy approximately this scaling while those close to the critical region (separating disorder from lane formation) exhibit a much slower asymptotic decay. In consequence, $g(r)$ cannot be easily used to detect or identify the particular state (\textit{disordered} or \textit{lanes}) in the system.

%

\begin{figure}[h!]
  \centering
  \includegraphics[width=0.65\linewidth]{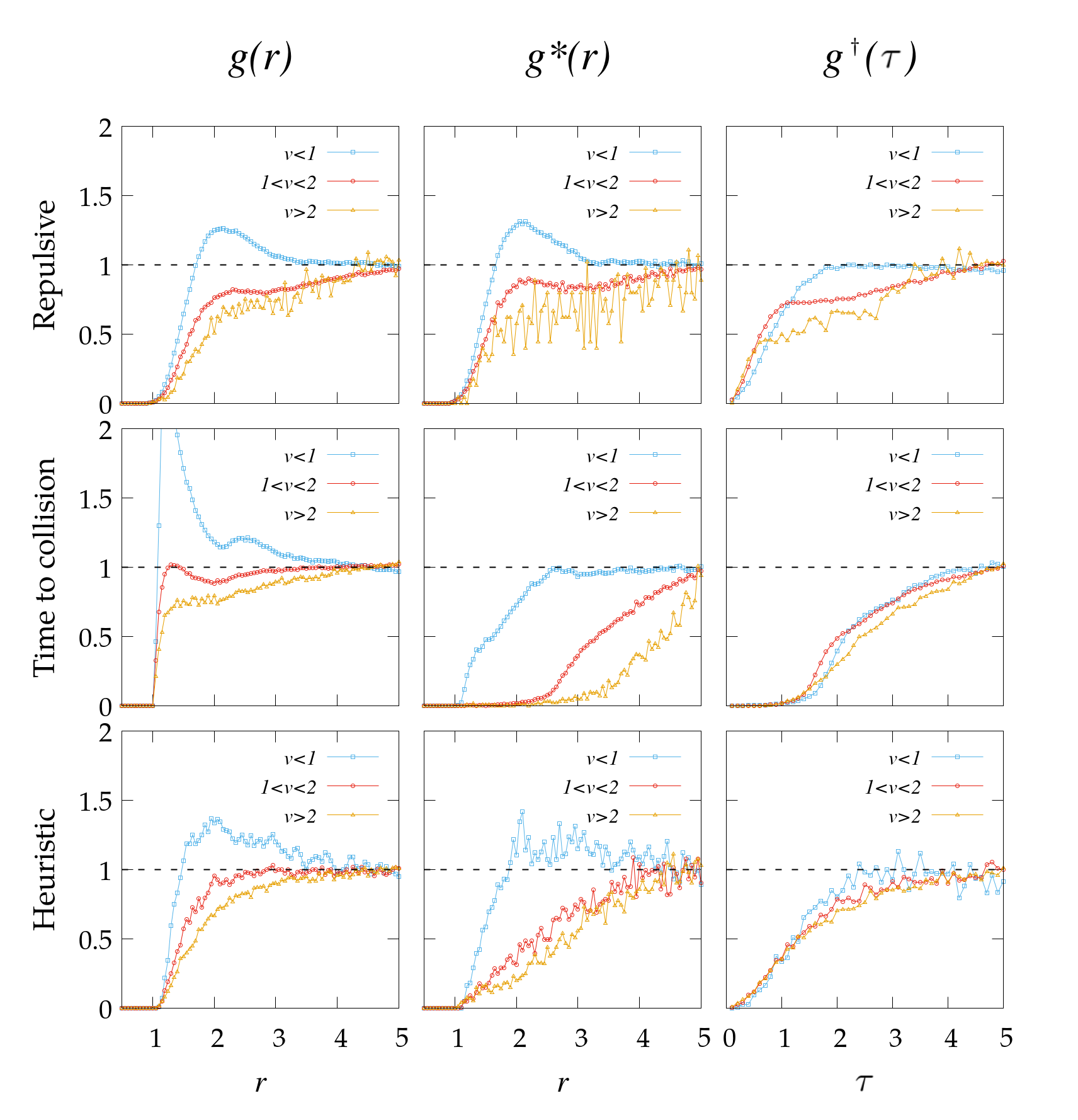}
    \caption{\textit{Radial and partial distribution functions $g(r)$, $g^{\ast}(r)$ and $g^{\dag}(\tau)$ when split into different regimes according to the relative speed between pairs $v_{r}$. Results are shown for $\rho=0.14$ and $\xi=0.1$, which corresponds to a disordered state (equivalent results for the phase with lanes are presented in the Supplementary Information for the sake of completeness). The repulsive interaction is fixed to $k=4$.}}
  \label{fig:rdfsep}
\end{figure}

Going further, we reproduce the analysis in \cite{karamouzas14} by showing how $g(r)$ gets modified if the data is split into three parts according to the relative speed between pairs of individuals $i$ and $j$, $v_r=\vert \vec{v}_i-\vec{v}_j \vert$. We find that individuals approaching each other with slow ($v_r<1$), intermediate ($1<v_r<2$) or fast ($v_r>2$) relative speeds exhibit very different behaviors in all cases (Fig. \ref{fig:rdfsep}, left column). However, we stress that, at least for the repulsive potential $F_{rep}^{sa}$ for which interactions are not velocity-dependent, the function $g(r)$ should be also independent of the actual velocity of the particles; this is confirmed in our simulations (see \textit{Suplemmentary Information file}) for consistency. 

The authors in \cite{karamouzas14} concluded that the differences observed in $g(r)$ for different values fo $v_r$, reflects that such function is not a very appropriate descriptor for capturing the effective interactions within the crowd or, stated in different words, the collective statistics of the system do not apparently yield a common behavior in the physical $r$ space of the distances between individuals. 
To explore here this idea we introduce a new magnitude $g^{*}(r)$, defined as the radial structure function but only for pairs of colliding agents, which are those for which $\tau$ is finite at that time step (Fig. \ref{fig:radialscheme}). 
The splitting of $g^{*}(r)$ into different values of the relative velocities still shows that the results are strongly dependent on $v_{r}$ (Fig. \ref{fig:rdfsep}, middle column), albeit the differences get reduced for $F_{\text{ttc}}^{(sa)}$ and $F_{\text{heu}}^{(sa)}$ (since these two interaction rules only apply to particles which are about to collide). Instead, for the rule $F_{\text{rep}}^{(sa)}$, which applies to all pairs of particles, the results found are almost the same as those for $g(r)$. So, again collective statistics seem to depart from such descriptor.


Our intuition, however, gained from the results in \cite{karamouzas14} is that the dynamics of self-avoidance should rather translate into a robust behavior within the $\tau$-space, as the events with low $\tau$ are the ones which must be avoided first. So, we finally introduce $g^{\dag}(\tau)$, which is the equivalent to $g(r)$ but on $\tau$-space, i.e. the density of agents found at a time-to-collision $\tau$ divided by the density we would find at the same $\tau$ for the case of non-interacting agents. The corresponding results are shown in Fig. \ref{fig:rdfsep} (right column). The idea that interactions should occur in the $\tau$-space is of course introduced by hand in our rule $F_{\text{ttc}}^{(sa)}$, and also implicitly in the rule $F_{\text{heu}}^{(sa)}$, so when we explore the dynamics in the $\tau$-space then we observe that the collapse between the three curves (for low, intermediate and high relative speeds) is almost perfect. However, we unexpectedly find that the collapse between the curves is moderately improved for $F_{\text{rep}}^{(sa)}$ too, though this self-avoiding rule has nothing to do with $\tau$. This suggests the existence of an underlying phenomena enhancing the relevance (at least at the level of how collective structures emerge) of the $\tau$-space whenever self-avoidance and bidirectionality effects drive the system dynamics. In the Supplementary Material  we carry out an alternative analysis of the radial distribution functions in order to provide additional support for this idea. Note that our results in Fig. \ref{fig:rdfsep} do not necessarily mean that pair interaction occurs in the $\tau$-space (which is not the case for our repulsive potential, actually) but that at a collective level this is the \textit{effective} situation produced.

Next step is to derive an effective potential of interaction between agents in the $\tau$-space. For the classical theory of fluids, the \textit{reversible work theorem} \cite{chandler87} in the $r$ space links the radial distribution function $g(r)$ with interaction energy between pairs in the form $V(r) \propto \ln [g(r)]$. Using an analogy with this classical result, the $\tau$-space also admits an equivalent derivation (see Appendix B). Such derivation is based on the idea that the system satisfies in the $\tau$-space a Boltzmann-like statistics. This cannot be justified from classical statistical mechanics since the concept of thermal equilibrium is meaningless in our context, but one can still invoke the Maximum Entropy Principle, which has solid foundations from information theory, to justify it at a statistical level \cite{jaynes57}. So that, the corresponding expression $V(\tau) \propto \ln [g^{\dag}(\tau)]$ derived in Appendix B must be interpreted as a statistical relation describing average properties in the $\tau$-space; note, however, that $V(\tau)$ does not represent the real potential of interaction between particles, and so $\nabla_{\mathbf{r}} V(\tau)$ is not to be interpreted as a physical force. 

\begin{figure}[h!]
  \centering
  \hspace{-3em}\includegraphics[width=0.65\linewidth]{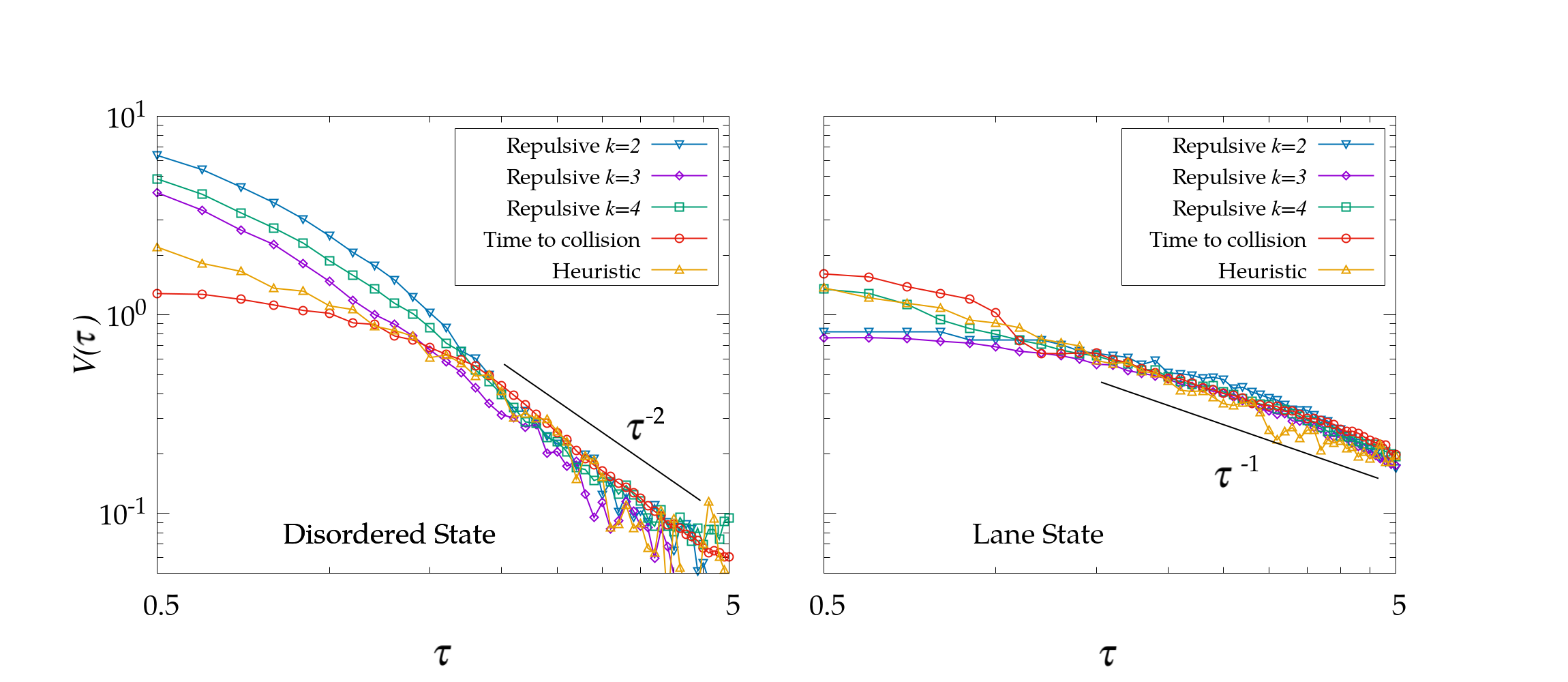}
    \caption{\textit{Effective interaction $V(\tau)$ obtained from $g^{\dag}(\tau)$ (see Eq. (\ref{rwt})) for $\rho=0.14$ in the different states of the phase space (the top image for $\xi=0.025$ and the bottom image for $\xi=2$).}}
  \label{fig:ln_gtau}
\end{figure}

The effective potential $V(\tau)$ obtained from our model is presented in Fig. \ref{fig:ln_gtau}, which represents the main result of our work. Surprisingly, we find that the three self avoiding mechanisms collapse for intermediate times-to collision (which is the significant region where most of the pair-pair interactions occur) into a common power-law relationship $V(\tau) \propto  \tau^{-\gamma}$, with $\gamma \approx 2$ for the disordered state and $\gamma \approx 1$ for the state with lanes (in Table 1 we show the results obtained from fitting the curves presented in Fig. \ref{fig:ln_gtau}). This common scaling is then apparently independent of the self-avoiding mechanism, and would be a direct consequence of the bidirectional nature of the flow considered. Note also that, contrary to what happened in the $r$ space (Fig. \ref{fig:OZ}), the $g^{\dag}(\tau)$ and the corresponding $V(\tau)$ show a different decay for the disordered phase and the case with lanes, so $g^{\dag}(\tau)$ can be effectively used to identify these two states.

According to Fig. \ref{fig:ang} (left), in the disordered state collisions can be produced in any orientation and the events corresponding to large $\tau$ are supressed by the shielding of closer events. The decay exponent $\gamma$ takes then a value of $2$, in agreement with the power-law proposed for pedestrians in \cite{karamouzas14}. While this is to be expected in the time-to-collision interaction $F_{\text{ttc}}^{(sa)}$ by definition, there is no apparent reason to justify why the same behavior emerges for $F_{\text{rep}}^{(sa)}$ and $F_{\text{heu}}^{(sa)}$. On the other side, the effective potential in the lane state exhibits a completely different behavior. The homogeneity in $\theta$ is there completely broken (Fig. \ref{fig:ang}, right) due to the two preferred directions of movement and most of the collisions are produced in the frontiers between opposite lanes. There is not shielding effect in this case and, as a consequence, the system is driven by a slower interaction decay, with $\gamma \approx 1$. We have not been able, however, to find an analytical justification for the specific values of $\gamma$ emerging in each state; this remains an open question.

\begin{table}[htpb!]
\centering
\begin{tabular}{ | c | c | c | }
\hline
  \&  $\gamma$ (disordered) \& $\gamma$ (lanes)  \\
  \hline
Repulsive $k=2$ \&  $2.2 \pm 0.3$ \&   $1.03 \pm 0.06$  \\
Repulsive $k=3$  \& $2.13 \pm 0.09$ \&  $1.01 \pm 0.04$  \\
Repulsive $k=4$   \& $2.07 \pm 0.09$ \& $0.99 \pm 0.05$   \\
Time to coll  \& $2.09 \pm 0.12$ \&  $1.08 \pm 0.07$  \\
Heuristic  \& $1.97 \pm 0.09$ \&  $1.04 \pm 0.06$  \\
\hline
\end{tabular}
\caption{\textit{Fit for the power law effective interaction $V(\tau)$ obtained from $g^{\dag}(\tau)$ (see Eq. (\ref{rwt})) for $\rho=0.14$ in the different states of the phase space (the left column for $\xi=0.025$ and the right column for $\xi=2$).}}
  \label{table:fit}
\end{table}

To conclude, from our findings in Fig. \ref{fig:ln_gtau} we obtain that the same effective potential, if computed through Eq. (\ref{rwt}) as done here, could emerge from a very wide range of interactions between the agents. This suggests that the result $V(\tau) \sim \tau^{-2}$ experimentally reported in \cite{karamouzas14} is not necessarily determining the actual rule of interaction (or self-avoidance) used by pedestrians, but it could rather be the manifestation of a common dynamics exhibited by a wide range of systems combining self-avoidance and bidirectionality. In particular, our results in Fig. 5 confirm that it is not possible to discern whether pedestrians use a time-to-colision potential (as in \cite{karamouzas14}) or a heuristic rule of path maximization (as in \cite{moussaid11}) only from examining the shape of the distribution function $g(\tau)$, but additional analysis would be required. Still, the scaling $V(\tau) \sim \tau^{-\gamma}$ will presumably work as a useful effective rule in bidirectional flows for different situations of interest (either pedestrian movement, ant organization or complex plasma \cite{morfil2009}, to cite some known cases). Such effective rule could be of great utility in order to simulate bidirectional fluxes without caring too much about the fine details of the interactions, and so it can be used as a toy or reference approximation to computational or analytical approaches in the field.

Our work is not able to delimit what is the exact range of validity of the scaling reported in Fig. 5. Given the strong differences between the three self-avoiding mechanisms explored here, we suspect that this range can be quite large as long as the initial scheme in Eq. (\ref{eq1}) holds. However, this does not guarantee the existence of an equivalent effective potential in situations where additional forces get introduced. Hence the extension of our results in this direction can represent an stimulating area of future research.

\section*{Appendix A: Methods}

\subsection*{i) Self-avoidance mechanisms}

As we explained before, we propose in our work three very different mechanism for the self-avoidance potential $\mathbf{F}^{(sa)}$ between the agents (which are considered as disk-like particles with the same diameter $D$). Here we provide a short description of each mechanism.

\label{sec:citeref} \textbf{Repulsive potential.}
We introduce a repulsive pair energy which is a function of the distance $r$ between the agents,

\begin{equation}
\mathbf{F}_{rep}^{(sa)} = -\frac{A}{r^{k}} \mathbf{u}
\label{E_lj}
\end{equation}
where $A>0$, $\mathbf{u}$ is a unit vector in the direction joining the pair of interacting agents, and the distance $r$ is measured in units of $D$, so $r=1$ corresponds to the distance between two adjacent agents. The parameter $k$ regulates the decay of the force, so implicitly it determines the range of scales where its effect is relevant, with the limit $k \to \infty$, reproducing hard-disk interactions.

This approach represents a reference model against which we subsequently compare the performance of more sophisticated mechanisms of self-avoidance between pedestrians. In this case the effect of $\mathbf{F}_{rep}^{(sa)}$ is to pull all the individuals apart, independently if their are moving forward to a collision or not.

\vspace{-0em}
\label{sec:citeref} \textbf{Time to collision.}
The second potential used corresponds to an interaction explicitly occurring in the time-to-collision (or $\tau$) space, defined as the time needed for the pair of agents to get in contact provided they followed their actual direction of motion. This time can be explicitly defined in terms of the relative velocity $\mathbf{v}_r$ and the relative position $\mathbf{r}$ between two given particles, 

\begin{equation}
\tau = \frac{-\mathbf{r}\mathbf{v}_{r}-\sqrt{(\mathbf{r}\mathbf{v}_{r})^{2} -|\mathbf{v}_{r}|^{2}(|\mathbf{r}|^{2}-D^{2})}}{|\mathbf{v}_{r}|^{2}}.
\label{eqtau}
\end{equation}

The idea of using the $\tau$-space for driving interactions is directly borrowed from \cite{karamouzas14} and is justified from the empirical results on pedestrians dynamics therein obtained. The rule reads then
 
\begin{equation}
\mathbf{F}_{tcc}^{(sa)} = - \nabla (k \tau^{-2} e^{-\frac{\tau}{\tau_{0}}}) ,
\label{E_time}
\end{equation}
with $k>0$, $\tau _0 >0$, and $u$ defined again as in (\ref{E_lj}). The exponential term is used as a cutoff to impair the effect of outermost collisions, so introducing the idea that agents possess a characteristic radius of perception ($\tau _0$, defined in the $\tau$-space). The potential so defined only applies to agents moving forward to a collision, such that $\tau$ can be defined and is positive; this is, pairs for which a positive value of $\tau$ cannot be found are considered as noninteracting agents. As a result, the set of agents interacting with a given one is a dynamic object which is updated continuously throughout the simulation.

\label{sec:citeref} \textbf{Heuristic rule.}
In order to consider very disparate mechanisms, we finally introduce a self-avoiding heuristic rule proposed for pedestrians dynamics in \cite{moussaid11}. This algorithm relies on the idea of adapting the direction of motion by maximizing locally the accessible distance path. So, each  agent samples its possible future trajectories by simulating internally (with a time horizon $t_m$) where it will reach by moving in a given direction (characterized by an angle $\alpha$) for some fixed time, provided that the other agents are assumed to go on moving in the same direction they do have at present. After sampling for a range of values of $\alpha$ (up to maximum $\alpha_{max}$, to avoid sudden or extreme changes of direction) the agent will choose the one that maximizes the length covered. After the election, all the agents reorient synchronously and the internal simulation starts anew.

There is a second rule, which determines the walking speed modulus after the reorientation. This is introduced in order to maintain a certain time to collision between the agent and the obstacle in the chosen walking direction \cite{moussaid11}. For this, we define a minimum time $\tau_{m}$ such that times-to-collision are forced to stay always below $\tau_m$ by reducing adequately the speed of the agents. That speed is then computed at practice as $v(t)=\text{min}[v,\frac{d_{obs}}{\tau_{m}}]$ where $d_{obs}$ is the distance between the agent and the first obstacle in the desired direction $\alpha$ at that time step.

\subsection*{ii) Implementation and technical details}
In this section we provide additional details of how the simulations of our multiagent model have been carried out.  For the first two interactions, the number of agents is fixed to $N=512$, while for the heuristic rule the number is fixed to $N=128$ due to the computational cost of its simulations (different time steps $\Delta t$ are also used in each case for the same reason, see below). The simulation time for the repulsive and time-to-collision mechanisms (it is, $F_{rep}^{(sa)}$ and $F_{ttc}^{(sa)}$) scales as $\propto N^{2}$, as they are pair to pair interactions. Instead, the heuristic rule prospects into the future the different $\alpha$ paths. This algorithm implies a scaling time $t \propto N^{2} m_{\alpha} d_{m}$, where $m_{\alpha}$ is the number of explored directions $\alpha$ in each evaluation of the rule (fixed in our case to $m_{\alpha}=50$), and $d_{m}=t_{m}/ \Delta t$ is the number of time steps in the prospection. Additionally, the simulation time step $\Delta t$ is also different in each one.

The agents are placed in a two-dimensional simulation box with density $\rho$ using periodic boundary conditions. The units of length are re-scaled to $\sigma$, so $r=1$ is equal to a diameter agent. The agent mass is settled as $m=1$. The Verlet algorithm has been used to integrate the equations.
The system is studied for different values of the density in the range $\rho=[0.05,0.32]$, which is accomplished by fixing the number of individuals to a certain value $N$ and changing the box size $L$, given $\rho=N/L^{2}$).

The parameters used for the implementation of the self-avoidance mechanisms are as follows: 

\begin{itemize}
\item The \textbf{repulsive interaction} (\ref{E_lj}) is fixed to $k=4$ (unless indicated otherwise), $A=2.5$ and $\Delta t_{rep}=0.001$.
\item The \textbf{time-to-collision potential} (\ref{E_time}) is fixed to  $k=1.5$, $\tau_{0}=10$ and $\Delta t_{ttc}=0.005$ according to \cite{karamouzas14}. The $\tau_{0}$ value is defined in order to not affect the dynamics in the scaling region. 
\item The \textbf{heuristic rule} is fixed to  $\tau_{m}=0.5$ and $d_{max}=v_{i}t_{m}$, with $t_{m}=5$,  $\alpha_{max}=75^\circ$ and $\Delta t_{heu}=0.05$, according to \cite{moussaid11}. 
\end{itemize}

To carry out the simulations 8 CPUs have been used, with a total simulation time around $\sim 250$ hours for each CPU.

\section{Appendix B: justification of the effective potential}
Here we provide a formal derivation of the relation between the distribution function $g^{\dagger}(\tau)$ and the effective potential $V(\tau)$. For classical physical systems, the reversible work theorem \cite{chandler87} provides such a formal relation between the classical radial distribution function $g(r)$ and the pair interaction potential $w(r)$, given by $ g(r)=\exp (-\beta w(r))$, by averaging the pairwise forces between two particles over the canonical ensemble.
Our derivation is a formal adaptation of the classical one to the case where the phase space is assumed to be defined by the times-to-collision between pairs of individuals, in agreement with the ideas stated throughout our article. Since our derivation works in the $\tau$-space then the concepts of force, work and potential must necessarily be interpreted in an effective (non-physical) way, as we shall see. The concept of thermal equilibrium is meaningless within this context. However, the idea of a canonical (Boltzmann-like) statistics is still attainable using an information-theory perspective in virtue of the Maximum Entropy principle (MEP), according to \cite{jaynes57,johnson80}. In particular, if we assume that our knowledge about the system is reduced to the average of an effective potential $V(\tau)$ in the $\tau$ phase space then the MEP yields immediately such a Boltzmann-like statistics.

Within this context, our derivation works as follows. Consider a system of $N$ particles where $V(\tau_{i,j})$ represents an effective pair interaction between individuals $i$ and $j$, with $\tau_{i,j}$ the time-to-collision between them. The global effective potential in the system reads then $V_N=V_N(\tau_{1,2}, \tau_{1,3}, \ldots,\tau_{N-1,N})= \sum_{i \neq j} V(\tau_{i,j})$. The corresponding phase space then consists of the $N(N-1)/2$ times-to-collision resulting from all possible pair interactions. We introduce now the magnitude $\nabla _{\tau_{1,2}} V_N$; for a classical conservative potential in the $\vec{r}$-space this would correspond to the force between particles $1$ and $2$, so one could be tempted to denote this magnitude as a generalized force. However, we will rather avoid such notation (i) to avoid misunderstandings coming from comparing our derivation to the classical one, and (ii) because that magnitude do not have dimensions of force, actually. If we average this magnitude in the Boltzmann-like statistics over the rest of coordinates of the phase space (this is, all except $\tau_{1,2}$) we have

\begin{equation}
\big \langle \nabla _{\tau_{1,2}} V_N \big \rangle  = \frac{ \int \left( \nabla _{\tau_{1,2}} V_N \right) e^{-\beta V_N} d\tau_{1,3} d\tau_{1,4} \ldots d\tau_{N-1,N}} { \int e^{-\beta V_N} d\tau_{1,3}  d\tau_{1,4} \ldots d\tau_{N-1,N} }.
\end{equation}

This can be rewritten as

\begin{equation}
-\big \langle \nabla _{\tau_{1,2}} V_N \big \rangle = \frac{1}{\beta} \nabla _{\tau_{1,2}} \left( \ln \int e^{-\beta V_N} d\tau_{1,2}  d\tau_{1,3} \ldots d\tau_{N-1,N} \right).
\end{equation}

Next, if we define the distribution function $g^{\dagger}(\tau)$ as the probability that the individuals $1$ and $2$ will be found to have a particular value $\tau_{1,2} = \tau$ then we expect within our Boltzmann-like scheme that

\begin{equation}
g^{\dagger}(\tau) \sim \int e^{-\beta V_N} d\tau_{1,3}  d\tau_{1,4} \ldots d\tau_{N-1,N}
\label{gV}
\end{equation} 
is satisfied. Accordingly, we can write

\begin{equation}
-\big \langle  \nabla _{\tau_{1,2}} V_N \big \rangle  = \frac{1}{\beta}  \nabla _{\tau_{1,2}} \left[ \ln \left( g^{\dagger}(\tau) \right) \right],
\label{prev}
\end{equation}
which is valid independently of the specific value of the normalization constant implicit in (\ref{gV}), since that constant is independent of $\tau_{1,2}$. Next, we observe that an effective pair potential $V(\tau)$ between particles 1 and 2 can be introduced as

\begin{equation}
V(\tau) \equiv \int^{0}_{\tau} \big \langle  \nabla _{\tau_{1,2}} V \big \rangle  d \tau_{1,2}
\end{equation}

Putting this expression together with (\ref{prev}) one obtains

\begin{equation}
V_{\tau} = - \frac{1}{\beta} \ln \left( g^{\dagger}(\tau) \right)
\label{rwt}
\end{equation}
after replacing the integration limits, and considering $V(\tau=0)=0$ to ensure that the behavior of the effective potential is meaningful in the $\tau$-space. 

Again, we warn that the derivation we have presented holds in a non-standard phase space and by using Boltzmann statistics from an information-theory perspective, without introducing any reference to thermal equilibrium in the classical sense. So that, our effective potential must really be interpreted as an effective magnitude describing how statistics work in the $\tau$-space, and therefore one is not allowed to interpret the spatial gradient of $V(\tau)$ as a real physical force between particles.

\bibliography{sample}

\section*{Acknowledgments}

This research has been supported by the Spanish government through Grants No. CGL2016-78156-C2-2-R and FIS2015-72434-EXP.
\section*{Author contributions statement}

 J.C., V.C. and D.C. conceived and designed the study. J.C. performed the numerical simulations. J.C., V.C. and D.C. carried out the mathematical analysis. J.C., V.C. and D.C. wrote and reviewed the paper.  The authors declare no competing interests.

\end{document}



\title{General scaling in bidirectional flows of self-avoiding agents \\ Supplementary Information \\}
\author{Javier Crist\'{i}n, Vicen\c{c} M\'{e}ndez and Daniel Campos \\
Grup de F\'{i}sica Estad\'{i}stica. Dept. de Fisica. Universitat Autonoma de Barcelona \\
08193 Bellaterra (Barcelona) Spain
}%

\date{\today}

\maketitle



In this Supplemental Material  we display further evidence for the phase transition and the universal interaction explained in the main text. 

\section*{Distribution function}
\begin{figure}[h]
  \centering
  \includegraphics[width=0.6\linewidth]{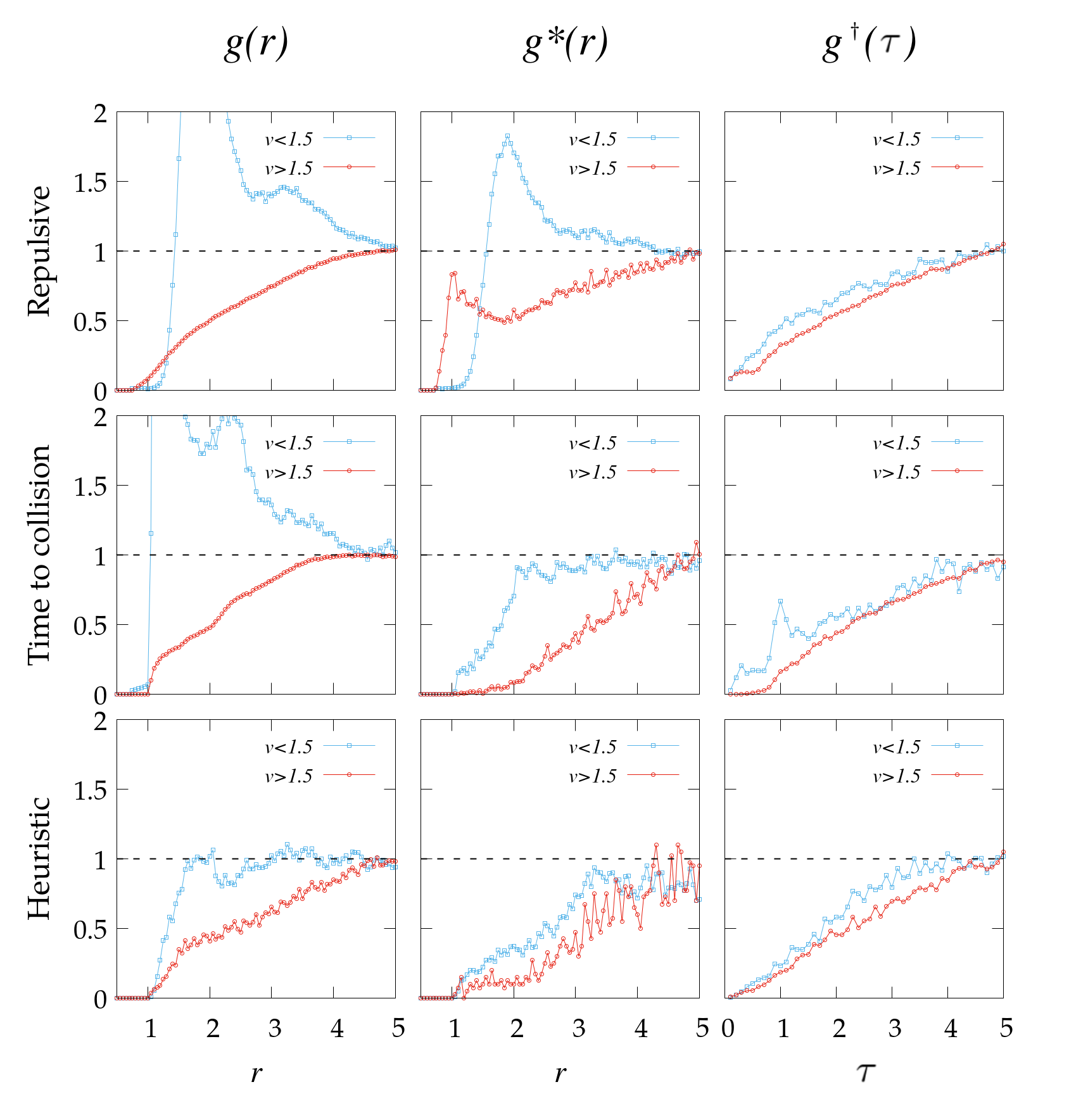}
    \caption{\textit{Radial and partial distribution functions $g(r)$, $g^{\ast}(r)$ and $g^{\dag}(\tau)$ when split into different regimes according to the relative speed between pairs $v_{r}$. Results are shown for $\rho=0.14$ and $\xi=4$, which corresponds to a lane state. The repulsive interaction is fixed to $k=4$}}
  \label{fig:rdfseplane}
\end{figure}

For the sake of completeness, we show at the Fig. \ref{fig:rdfseplane} the equivalent to the partial distribution analysis (main text, Fig. 4) but now for the lanes state. The splitting in different ranges of $v_{r}$ in this case has to be necessarily different to that for the disordered state due to the lack of intermediate relative velocities when the lanes are formed. All agents will be moving either leftwards or rightwards (according to the preferred direction of each). Then relative velocities are either very low ($v_r<1.5$, for agents in the same lane) or very large ($v_r>1.5$, for agents in lanes with opposite directions), while there is no statistics at the intermediate regime $1<v_r<2$ used in Figure 4 in the main text. Additionally, we often observe that particles persistently moving in the same lane and very close to each other (so, with $\tau \rightarrow 0$) artificially dominate the statistics, so these have been removed when computing $g^{\dag}(\tau)$ in Fig. \ref{fig:rdfseplane}.

The  collapse observed in the distribution $g^{\dag}(\tau)$ for different $v_r$ ranges confirms the behavior already reported for the disordered phase (note that now even the repulsive case satisfies the collapse to a great extent). This further supports the existence of an underlying phenomena enhancing the prominence of the $\tau$-space whenever self-avoidance and bidirectionality effects drive the system dynamics. 

\section{Behaviour of the repulsive potential}

\begin{figure}[h]
	\centering
	\includegraphics[width=0.6\linewidth]{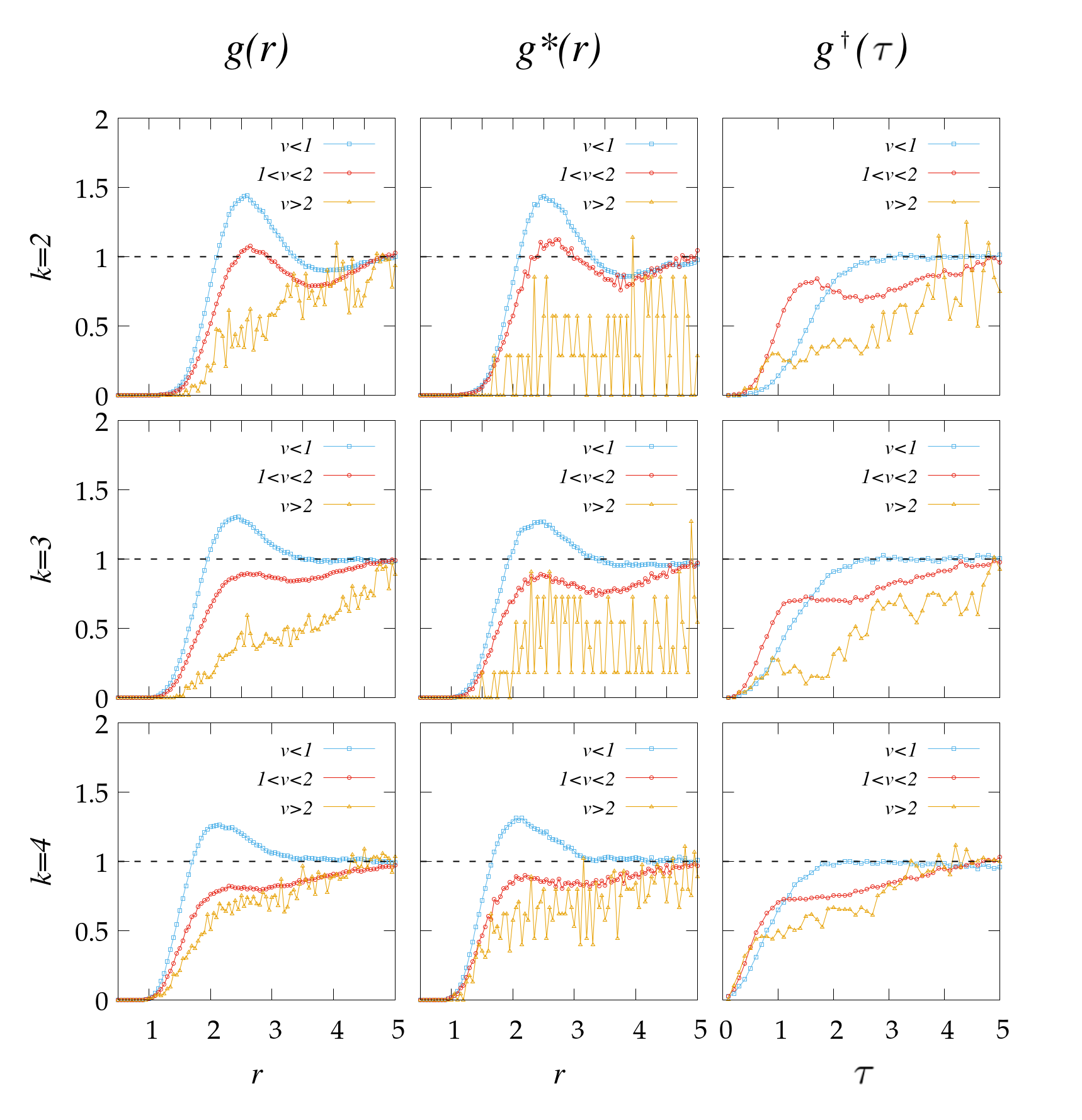}
	\caption{\textit{Distribution functions $g(r)$, $g^{\ast}(r)$ and $g^{\dag}(\tau)$ when split into different regimes according to the relative speed between pairs $v_{r}$. The example system is formed by agents interacting with the  repulsive potential with a) $k=2$ b) $k=3$  c) $k=4$ at the phase space point $\rho=0.14$ and $\xi=0.1$.}}
	\label{fig:rdfsep}
\end{figure}

While in the Figure 4 of the main text we present the results for the repulsive potential with $k=4$ for the sake of simplicity, we here complete the comparison study by showing that other values of $k$ lead to very similar results (in particular the cases $k=2$ and $k=3$ are shown). We compare the distribution functions $g(r)$, $g^{\ast}(r)$ and $g^{\dag}(\tau)$ (Fig. \ref{fig:rdfsep}) when they are split according to the different values of the relative speed $v_{r}$ (Fig. \ref{fig:rdfsep}). In agreement with our discussion in the main text, we find that the curves tend to converge in the case where the $\tau$-space is considered. Although the convergence is far from being perfect (as is to be expected, actually, since $F_{rep}^{(sa)}$ is not defined in the $\tau$-space), we still find that the effective interaction potential, defined through $V(\tau) \sim \ln \left[ g^{\dag}(\tau) \right]$, still satisfies the same scaling $V(\tau) \sim \tau ^{\gamma}$ with $\gamma \approx 2$ in the disordered state and $\gamma \approx 1$ for lanes. As a whole, this analysis confirm the universal nature of the scaling reported for $V(\tau)$, which seems to be independent of the self-avoiding mechanism considered, even for those cases where this mechanism has nothing to do originally with the $\tau$-space.

In the Figure 4 of the main text we have already proved that $g(r)$ exhibits very different shapes when data is split according to the relative speed between the pairs of particles interacting. However, at least for the case of the repulsive potential, where the force between pairs $\mathbf{F}^{sa}$ (and so the corresponding potential resulting from the integration of $\mathbf{F}$) depend only on the distance $r$, we must expect that $g(r)$ is well-defined (since in this case the classical reversible work theorem holds). Then a real (physical) potential $V_p(r)$ can be properly defined (but this does not necessarily have any formal relation with the effective potential $V(\tau)$ analyzed in the main work). To verify that our system is well-behaved then we include here the behavior obtained for the repulsive case $F_{rep}^{sa}$ when $g(r)$ is split according to the actual value of the particles speed $v$ (and not $v_r$, as in Figure 4 of the main text). The overlap between all the curves (independent of $v$) confirms that our simulated system is physically sound, and so supports the idea that the general behavior reported for $V(\tau)$ is not an artifact.

\begin{figure}[h]
	\centering
	\includegraphics[width=0.9\linewidth]{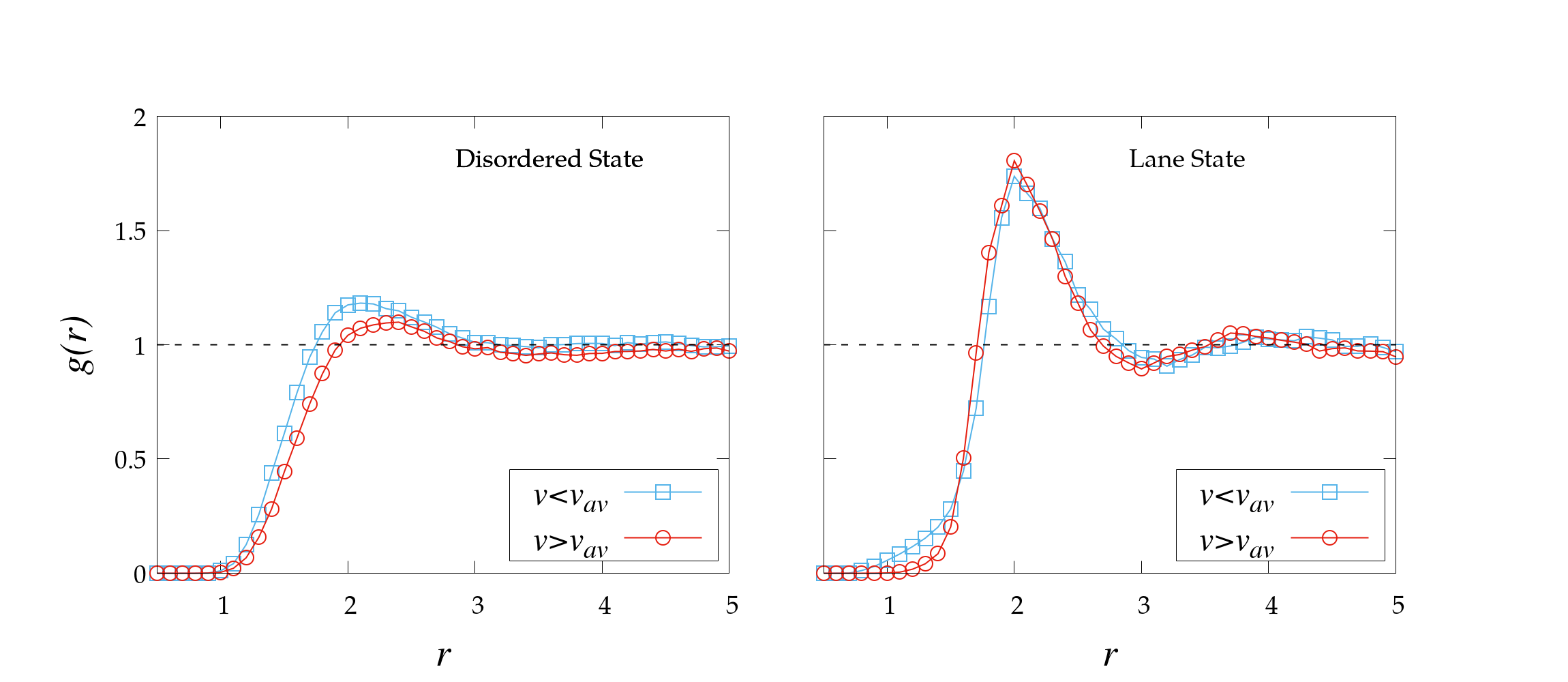}
	\caption{\textit{Radial distribution function $g(r)$ when split into two different regimes according to the particle speed $v$. The left image corresponds to $\xi=0.1$ and the right image for $\xi=2$, with $\rho=0.14$ and $k=4$ in both cases.}}
	\label{fig:indvel}
\end{figure}